\documentstyle[aps]{revtex}

\begin{document}
\title{Massive Gauge Field Theory Without Higgs Mechanism\\
I. Ward-Takahashi Identities and Proof of Unitarity}
\author{Jun-Chen Su}
\address{Center for Theoretical Physics,College of Physics, Jilin University,\\
Changchun 130023,\\
People's Republic of China}
\date{}
\maketitle

\begin{abstract}
In our previously published papers, it was argued that a massive non-Abelian
gauge field theory in which all gauge fields have the same mass can well be
set up on the gauge-invariance principle. The quantization of the fields was
performed by different methods. In this paper, It is proved that the quantum
theory is invariant with respect to a kind of BRST-transformations. From the
BRST-invariance of the theory , the Ward-Takahashi identities satisfied by
the generating functionals of full Green's functions, connected Green's
functions and proper vertex functions are successively derived. As an
application of the above Ward-Takahashi identity, the Ward-Takahashi
identity obeyed by the massive gauge boson propagator is derived and the
renormalization of the propagator is discussed. Furthermore, based on the
Ward-Takahashi identity, it is exactly proved that the S-matrix elements
given by the quantum theory are gauge-independent and hence unitary.

\end{abstract}

\section{Introduction}

In our preprevious papers$^{[1-4]}$, it has been shown that some massive
gauge field theory in which the masses of all gauge fields are the same may
really be set up on the gauge-invariance principle without the help of the
Higgs mechanism. The essential points of achieving this conclusion are the
following. (1) The massive gauge field must be viewed as a constrained
system in the whole space of vector potential. Therefore, the Lorentz
condition, as a necessary constraint, must be introduced from the onset and
imposed on the massive Yang-Mills Lagrangian so as to restrict the
unphysical degrees of freedom involved in the Lagrangian; (2) The
gauge-invariance of gauge field dynamics should be more generally required
to the action of the field other than the Lagrangian because the action is
of more fundamental dynamical meaning. In particular, the gauge-invariance
for the constrained system should be required to the action written in the
physical subspace defined by the Lorentz condition in which the fields exist
and move only; (3) In the physical subspace , only the infinitesimal gauge
transformations are possible to exist and necessary to be considered in
inspection of whether the theory is gauge-invariant or not; (4) To construct
a correct gauge field theory, the residual gauge degrees of freedom existing
in the physical subspace must be eliminated by the constraint condition on
the gauge group. This constraint condition may be determined by requiring
the action to be gauge-invariant. Thus, the theory is set up from beginning
to end on the gauge-invariance principle. These points are important to
build up a correct quantum massive non-Abelian gauge field theory. Such a
theory, as will be proved, is renormalizable and unitary.

In Refs. [1-3], the quantum theory of the massive non-Abelian gauge fields
without Higgs mechanism was established by different methods of
quantization. In this paper, it will be shown that the quantum theory has an
important property that the effective action appearing in the generating
functional of Green's functions is invariant with respect to a kind of
BRST-transformations$^{[5]}$. From the BRST-symmetry, we will derive various
Ward-Takahashi (W-T) identities$^{[6-12]}$ satisfied by the generating
functionals of Green's functions and proper vertices. These W-T identities
are of special importance in proofs of unitarity and renormalizability of
the theory. In this paper, we confine ourselves to prove that the S-matrix
elements evaluated from the theory is independent of the gauge parameter,
that is to say, the gauge-dependent unphysical poles appearing in the gauge
boson propagator and the ghost particle propagator do not contribute to the
S-matrix elements. Therefore, the unitarity of the theory is ensured.

The arrangement of this paper is as follows. In section 2, we will derive
the BRST-transformations under which the effective action of the massive
non-Abelian gauge field theory is invariant. In doing this, we extend our
discussion by including fermions. In section 3, we will derive the W-T
identities satisfied by various generating functionals. In section 4, to
illustrate applications of the above W-T identity, the W-T identity obeyed
by the massive gluon propagator will be derived and the renormalization of
the propagator will be discussed. In section 5, by virtue of the W-T
identity, we will prove that the S-matrix elements calculated from the
massive gauge field theory are gauge-independent and hence unitary. The last
section is used to make some remarks on the nilpotency problem of the
BRST-transformations and the BRST-invariace of the external source terms
introduced to the generating functional. In Appendix, the W-T identities
used to prove the unitarity will be given by an alternative derivation.

\section{ BRST- transformation}

In the previous paper , we mainly discussed the gauge fields themselves
without concerning fermion fields. For the gauge fields, in order to
guarantee the mass term in the action to be gauge-invariant, the masses of
all gauge fields are taken to be the same. If fermions are included, as
pointed out in Refs.[1-3], the QCD with massive gluons fulfils this
requirement because all the gluons can be considered to have the same mass $%
m $. The SU(3)-symmetric action of the QCD with massive gluons is given by
the following Lagrangian$^{[1-4]}$ 
\begin{equation}
{\cal L}=\bar \psi \{i\gamma ^\mu (\partial _\mu -igT^aA_\mu ^a)-M\}\psi -%
\frac 14F^{a\mu \nu }F_{\mu \nu }^a+\frac 12m^2A^{a\mu }A_\mu ^a  \eqnum{2.1}
\end{equation}
where $\psi (x)$ denotes the quark field function , $\bar \psi (x)$ is its
Dirac-conjugate, $T^a=\lambda ^a/2$ are the color matrices and $M$ is the
quark mass. The above Lagrangian is constrained by the Lorentz condition 
\begin{equation}
\partial ^\mu A_\mu ^a=0  \eqnum{2.2}
\end{equation}
Under this condition, as was proved in Refs. [1-3], the action given by the
Lagrangian in Eq. (2.1) is invariant with respect to the following gauge
transformations: 
\begin{equation}
\begin{array}{c}
\delta A_\mu ^a=\xi D_\mu ^{ab}C^b \\ 
\delta \psi (x)=ig\xi T^aC^a(x)\psi (x) \\ 
\delta \bar \psi (x)=ig\xi \bar \psi (x)T^aC^a(x)
\end{array}
\eqnum{2.3}
\end{equation}
where we have set the parametric functions of the gauge group $\theta
^a(x)=\xi C^a(x)$ in which $\xi $ is an infinitesimal Grassmann number and $%
C^a(x)$ are the ghost field functions. According to the result given in
paper I, the quantum theory built up from the Lagrangian in Eq. (2.1) and
the constraint condition in Eq. (2.2) is described by the following
generating functional of Green's functions$^{[5-9]}$%
\begin{equation}
\begin{array}{c}
Z[J_\mu ^a,\bar K^a,K^a,\bar \eta ,\eta ]=\frac 1N\int {\cal D}(A_\mu ^a,%
\bar C^a,C^a,\bar \psi ,\psi )exp\{iS \\ 
+i\int d^4x[J^{a\mu }A_\mu ^a+\bar K^aC^a+\bar C^aK^a+\bar \psi \eta +\bar 
\eta \psi ]\}
\end{array}
\eqnum{2.4}
\end{equation}
where $J^{a\mu }$, $\bar K^a$, $K^a$, $\eta $ and $\bar \eta $ are the
external sources and 
\begin{equation}
\begin{array}{c}
S=\int d^4x\{\bar \psi [i\gamma ^\mu (\partial _\mu -igT^aA_\mu ^a)-M]\psi -%
\frac 14F^{a\mu \nu }F_{\mu \nu }^a+\frac 12m^2A^{a\mu }A_\mu ^a \\ 
-\frac 1{2\alpha }(\partial ^\mu A_\mu ^a)^2+\bar C^a\partial ^\mu ({\cal D}%
_\mu ^{ab}C^b)\}
\end{array}
\eqnum{2.5}
\end{equation}
is the effective action in which

\begin{equation}
{\cal D}_\mu ^{ab}(x)=\frac{\mu ^2}{\Box _x}\partial _\mu ^x+D_\mu ^{ab}(x) 
\eqnum{2.6}
\end{equation}
here $\mu ^2=\alpha m^2$ and 
\begin{equation}
D_\mu ^{ab}=\delta ^{ab}\partial _\mu -gf^{abc}A_\mu ^c  \eqnum{2.7}
\end{equation}
is the covariant derivative. Similar to the massless gauge theory, for the
massive gauge theory, there are a set of BRST-transformations including the
infinitesimal gauge transformations shown in Eq. (2.3) and the
transformations for the ghost fields under which the effective action is
invariant. The transformations for the ghost fields may be found from the
stationary condition of the effective action under the BRST-transformations.
By applying the transformations in Eq. (2.3) to the action in Eq. (2.5), one
can derive 
\begin{equation}
\delta S=\int d^4x\{[\delta \bar C^a-\frac \xi \alpha \partial ^\nu A_\nu
^a]\partial ^\mu ({\cal D}_\mu ^{ab}C^b)+\bar C^a\partial ^\mu \delta ({\cal %
D}_\mu ^{ab}C^b)\}=0  \eqnum{2.8}
\end{equation}
This expression suggests that if we set 
\begin{equation}
\delta \bar C^a=\frac \xi \alpha \partial ^\nu A_\nu ^a  \eqnum{2.9}
\end{equation}
and 
\begin{equation}
\partial ^\mu \delta ({\cal D}_\mu ^{ab}C^b)=0  \eqnum{2.10}
\end{equation}
The action will be invariant. Eq. (2.9) gives the transformation law of the
ghost field variable $\bar C^a(x)$ which is the same as the one in the
massless gauge field theory. From Eq. (2.10), we may derive a transformation
law of the ghost variables $C^a(x)$. Noticing the relation in Eq. (2.6), we
can write 
\begin{equation}
\delta ({\cal D}_\mu ^{ab}(x)C^b(x))=\frac{\mu ^2}{\Box _x}\partial _\mu
^x\delta C^a(x)+\delta (D_\mu ^{ab}(x)C^b(x))  \eqnum{2.11}
\end{equation}
In the massless gauge theory, it has been proved that$^{[8-11]}$

\begin{equation}
\delta (D_\mu ^{ab}(x)C^b(x))=D_\mu ^{ab}(x)[\delta C^b(x)+\frac \xi 2%
gf^{bcd}C^c(x)C^d(x)]  \eqnum{2.12}
\end{equation}
With this result, Eq. (2.11) can be written as 
\begin{equation}
\delta ({\cal D}_\mu ^{ab}(x)C^b(x))={\cal D}_\mu ^{ab}(x)\delta
C^b(x)-D_\mu ^{ab}(x)\delta C_0^b(x)  \eqnum{2.13}
\end{equation}
where 
\begin{equation}
\delta C_0^a(x)\equiv -\frac{\xi g}2f^{abc}C^b(x)C^c(x)  \eqnum{2.14}
\end{equation}
On substituting Eq. (2.13) into Eq. (2.10), we have 
\begin{equation}
M^{ab}(x)\delta C^b(x)=M_0^{ab}(x)\delta C_0^b(x)  \eqnum{2.15}
\end{equation}
where we have defined 
\begin{equation}
M^{ab}(x)\equiv \partial _x^\mu {\cal D}_\mu ^{ab}(x)=\delta ^{ab}(\Box
_x+\mu ^2)-gf^{abc}A_\mu ^c(x)\partial _x^\mu  \eqnum{2.16}
\end{equation}
and

\begin{equation}
M_0^{ab}(x)\equiv \partial _x^\mu D_\mu ^{ab}(x)=M^{ab}(x)-\mu ^2\delta ^{ab}
\eqnum{2.17}
\end{equation}
It is noted that the operator in Eq. (2.16 ) is just the operator appearing
in the equation of motion for the ghost field $C^a(x)$%
\begin{equation}
\partial ^\mu ({\cal D}_\mu ^{ab}C^b)=0  \eqnum{2.18}
\end{equation}
(see Eq. (4.10) in paper I). Corresponding to this equation of motion, we
may write an equation satisfied by the Green's function $\Delta ^{ab}(x-y)$ 
\begin{equation}
M^{ac}(x)\Delta ^{cb}(x-y)=\delta ^{ab}\delta ^4(x-y)  \eqnum{2.19}
\end{equation}
The function $\Delta ^{ab}(x-y)$ is nothing but the exact propagator of the
ghost field which is inverse of the operator $M^{ab}(x)$. In the light of
Eq. (2.19) and noticing Eq. (2.17), we may solve out the $\delta C^a(x)$
from Eq. (2.15) 
\begin{equation}
\begin{array}{c}
\delta C^a(x)=(M^{-1}M_0\delta C_0)^a(x)=\{M^{-1}(M-\mu ^2)\delta C_0\}^a(x)
\\ 
=\delta C_0^a(x)-\mu ^2\int d^4y\Delta ^{ab}(x-y)\delta C_0^b(y)
\end{array}
\eqnum{2.20}
\end{equation}
This just is the transformation law for the ghost variables $C^a(x)$. When
the mass tends to zero, Eq. (2.20) immediately goes over to the
corresponding transformation given in the massless gauge field theory. It is
interesting that in the Landau gauge ($\alpha =0),$ due to $\mu =0$, the
above transformation also reduces to the form as given in the massless
theory. This result is natural since in the Landau gauge, the gauge field
mass term in the action is gauge-invariant. However, in general gauges, the
mass term is no longer gauge-invariant. In this case, to maintain the action
to be gauge-invariant, it is necessary to give the ghost field a mass $\mu $
so as to counteract the gauge-non-invariance of the gauge field mass term.
As a result, in the transformation given in Eq. (2.20) appears a term
proportional to $\mu ^2.$

\section{Ward-Takahashi identities}

This section serves to derive the W-T identities for the quantum massive
non-Abelian gauge field theory established in paper I and represented in
Eqs. (2.4)-(2.7) on the basis of the BRST-symmetry of the theory. Since
derivations of the W-T identities for the QCD with massive gluons are much
similar to those for the QCD with massless gluons, we only need here to give
a brief description of the derivations. When we make the
BRST-transformations shown in Eqs. (2.3), (2.9) and (2.20) to the generating
functional in Eq.(2.4) and consider the invariance of the generating
functional, the action and the integration measure under the transformations
(the invariance of the integration measure is easy to check), we obtain an
identity such that 
\begin{equation}
\begin{array}{c}
\frac 1N\int {\cal D}(A_\mu ^a,\bar C^a,C^a,\bar \psi ,\psi )\int
d^4x\{J^{a\mu }(x)\delta A_\mu ^a(x)+\delta \overline{C}^a(x)K^a(x)+\bar K%
^a(x)\delta C^a(x) \\ 
+\bar \eta (x)\delta \psi (x)+\delta \bar \psi (x)\eta (x)\}e^{iS+EST} \\ 
=0
\end{array}
\eqnum{3.1}
\end{equation}
where $EST$ is an abbreviation of the external source terms appearing in Eq.
(2.4). The Grassmann number $\xi $ contained in the BRST-transformations in
Eq. (3.1) may be eliminated by performing a partial differentiation of Eq.
(3.1) with respect to $\xi $. As a result, we get a W-T identity as follows 
\begin{equation}
\begin{array}{c}
\frac 1N\int {\cal D}(A_\mu ^a,\bar C^a,C^a,\bar \psi ,\psi )\int
d^4x\{J^{a\mu }(x)\Delta A_\mu ^a(x)+\triangle \overline{C}^a(x)K^a(x)-\bar K%
^a(x)\Delta C^a(x) \\ 
-\bar \eta (x)\Delta \psi (x)+\Delta \bar \psi (x)\eta (x)\}e^{iS+EST} \\ 
=0
\end{array}
\eqnum{3.2}
\end{equation}
where 
\begin{equation}
\begin{array}{c}
{\Delta }A_\mu ^a(x)=D_\mu ^{ab}(x)C^b(x) \\ 
{\Delta }\bar C^a(x)=\frac 1\alpha \partial ^\mu A_\mu ^a(x) \\ 
{\Delta }C^a(x)=\int d^4y[\delta ^{ab}\delta ^4(x-y)-\mu ^2\Delta ^{ab}(x-y)]%
{\triangle }C_0^b(y) \\ 
{\Delta }C_0^b(y)=-\frac 12gf^{bcd}C^c(y)C^d(y) \\ 
{\Delta }\psi (x)=igT^aC^a(x)\psi (x) \\ 
{\Delta }\bar \psi (x)=ig\bar \psi (x)T^aC^a(x)
\end{array}
\eqnum{3.3}
\end{equation}
These functions defined above are finite. Each of them differs from the
corresponding BRST-transformation written in Eqs. (2.3), (2.9) and (2.20) by
an infinitesimal Grassmann parameter $\xi .$

In order to represent the composite field functions $\Delta A_\mu ^a,\Delta
C^a,\Delta \bar \psi $ and $\Delta \psi $ in Eq. (3.2) in terms of
differentials of the functional $Z$ with respect to external sources, we
may, as usual, construct a generalized generating functional by introducing
new external sources (called BRST-sources later on) into the generating
functional written in Eq. (2.4), as shown in the following$^{[8-11]}$%
\begin{equation}
\begin{array}{c}
Z[J_\mu ^a,\bar K^a,K^a,\bar \eta ,\eta ;u^{a\mu },v^a,\bar \zeta ,\zeta ]
\\ 
=\frac 1N\int {\cal D}[A_\mu ^a,\bar C^a,C^a,\bar \psi ,\psi ]exp\{iS+i\int
d^4x[u^{a\mu }\Delta A_\mu ^a+v^a\Delta C^a \\ 
+\Delta \bar \psi \zeta +\bar \zeta \Delta \psi +J^{a\mu }A_\mu ^a+\bar K%
^aC^a+\bar C^aK^a+\bar \eta \psi +\bar \psi \eta ]\}
\end{array}
\eqnum{3.4}
\end{equation}
where $u^{a\mu },$ $v^a$, $\overline{\varsigma }$ and $\varsigma $ are the
sources which belong to the corresponding functions $\Delta A_{\mu \text{ , }%
}^a\Delta C^a$, $\Delta \Psi $ and $\Delta \overline{\Psi \text{ }}$
respectively. Obviously, the $u^{a\mu }$ and $\Delta A_\mu ^a$ are
anticommuting quantities, while, the $v^a$, $\bar \zeta $, $\zeta $, $\Delta
C^a$, $\Delta \bar \psi $ and $\Delta \psi $ are commuting ones. We may
start from the above generating functional to re-derive the W-T identity. In
order that the identity thus derived is identical to that as given in Eq.
(3.2), it is necessary to require the BRST-source terms $u_i\Delta \Phi _i$
where $u_i=u^{a\mu }$, $v^a$, $\overline{\zeta }$ or $\zeta $ and $\Delta
\Phi _i=\Delta A_\mu ^a$, $\Delta C^a$, $\Delta \Psi $ or $\Delta \overline{%
\Psi \text{ }}$ to be invariant under the BRST-transformations. How to
ensure the BRST-invariance of the source terms? For illustration, let us
introduce the source terms in such a fashion 
\begin{equation}
\begin{array}{c}
\int d^4x[\widetilde{u}^{a\mu }\delta A_\mu ^a+\widetilde{v}^a\delta C^a+%
\overline{\widetilde{\zeta }}\delta \psi +\delta \overline{\psi }\widetilde{%
\zeta }] \\ 
=\int d^4x[u^{a\mu }\triangle A_\mu ^a+v^a\triangle C^a+\overline{\zeta }%
\triangle \psi +\triangle \overline{\psi }\zeta ]
\end{array}
\eqnum{3.5}
\end{equation}
where 
\begin{equation}
u^{a\mu }=\tilde u^{a\mu }\xi ,\;\;v^a=\tilde v^a\xi ,\;\;\bar \varsigma =%
\overline{\widetilde{\varsigma }}\xi ,\;\;\varsigma =-\tilde \varsigma \xi 
\eqnum{3.6}
\end{equation}
These external sources are defined by including the Grassmann number $\xi $
and hence products of them with $\xi $ vanish. This suggests that we may
generally define the sources by the following condition 
\begin{equation}
u_i\xi =0  \eqnum{3.7}
\end{equation}
Considering that under the BRST-transformation, the variation of the
composite field functions given in the general gauges can be represented in
the form $\delta \Delta \Phi _i=\xi \widetilde{\Phi }_i$ where $\widetilde{%
\Phi }_i$ are functions without including the parameter $\xi $, clearly, the
definition in Eq. (3.7) for the sources would guarantee the BRST- invariance
of the BRST-source terms. When the BRST-transformations in Eqs. (2.3), (2.9)
and (2.20) are made to the generating functional in Eq. (3.4), due to the
definition in Eq. (3.7) for the sources, we have $u_i\delta \Delta \Phi _i=0$
which means that the BRST-source terms give a vanishing contribution to the
identity in Eq. (3.1). Therefore, we still obtain the identity as shown in
Eq. (3.1) except that the external source terms is now extended to include
the BRST-external source terms. This fact indicates that we may directly
insert the BRST-source terms into the exponent in Eq. (3.1) without changing
the identity itself. When performing a partial differentiation of the
identity with respect to $\xi $, we obtain a W-T identity which is the same
as written in Eq. (3.2) except that the BRST-source terms are now included
in the identity. Therefore, Eq. (3.2) may be expressed as 
\begin{equation}
\begin{array}{c}
\int d^4x[J^{a\mu }(x)\frac \delta {\delta u^{a\mu }(x)}-\bar K^a(x)\frac 
\delta {\delta v^a(x)}-\bar \eta (x)\frac \delta {\delta \bar \zeta (x)} \\ 
+\eta (x)\frac \delta {\delta \zeta (x)}+\frac 1\alpha K^a(x)\partial _x^\mu 
\frac \delta {\delta J^{a\mu }(x)}]Z[J_\mu ^a,\cdots ,\zeta ] \\ 
=0
\end{array}
\eqnum{3.8}
\end{equation}
This is the W-T identity satisfied by the generating functional of full
Green functions.

On substituting in Eq. (3.8) the relation $^{6-8}$ 
\begin{equation}
Z=e^{iW}  \eqnum{3.9}
\end{equation}
where W denotes the generating functional of connected Green's functions,
one may obtain a W-T identity expressed by the functional W 
\begin{equation}
\begin{array}{c}
\int d^4x[J^{a\mu }(x)\frac \delta {\delta u^{a\mu }(x)}-\bar K^a(x)\frac 
\delta {\delta v^a(x)}-\bar \eta (x)\frac \delta {\delta \bar \zeta (x)}%
+\eta (x)\frac \delta {\delta \zeta (x)} \\ 
+\frac 1\alpha K^a(x)\partial _x^\mu \frac \delta {\delta J^{a\mu }(x)}%
]W[J_u^a,\cdots ,\zeta ] \\ 
=0
\end{array}
\eqnum{3.10}
\end{equation}
From this identity, one may get another W-T identity satisfied by the
generating functional $\Gamma $ of proper (one-particle-irreducible) vertex
functions. The functional $\Gamma $ is usually defined by the following
Legendre transformation$^{[8-11]}$%
\begin{equation}
\begin{array}{c}
\Gamma [A^{a\mu },\bar C^a,C^a,\bar \psi ,\psi ;u_\mu ^a,v^a,\bar \zeta
,\zeta ]=W[J_\mu ^a,\bar K^a,K^a,\bar \eta ,\eta ;u_\mu ^a,v^a,\bar \zeta
,\zeta ] \\ 
-\int d^4x[J_\mu ^aA^{a\mu }+\bar K^aC^a+\bar C^aK^a+\bar \eta \psi +\bar 
\psi \eta ]
\end{array}
\eqnum{3.11}
\end{equation}
where $A_\mu ^a,\bar C^a,C^a,\bar \psi $ and $\psi $ are the field variables
defined by the following functional derivatives$^{[5-8]}$%
\begin{equation}
\begin{array}{c}
A_\mu ^a(x)=\frac{\delta W}{\delta J^{a\mu }(x)},\;\;\bar C^a(x)=-\frac{%
\delta W}{\delta K^a(x)},C^a(x)=\frac{\delta W}{\delta \bar K^a(x)}, \\ 
\bar \psi (x)=-\frac{\delta W}{\delta \eta (x)},\;\;\psi (x)=\frac{\delta W}{%
\delta \bar \eta (x)}
\end{array}
\eqnum{3.12}
\end{equation}
From Eq.(3.11), it is not difficult to get the inverse transformations$%
^{[5-8]}$%
\begin{equation}
\begin{array}{c}
J^{a\mu }(x)=-\frac{\delta \Gamma }{\delta A_\mu ^a(x)},\;\;\bar K^a(x)=%
\frac{\delta \Gamma }{\delta C^a(x)},K^a(x)=-\frac{\delta \Gamma }{\delta 
\bar C^a(x)}, \\ 
\bar \eta (x)=\frac{\delta \Gamma }{\delta \psi (x)},\;\;\eta (x)=-\frac{%
\delta \Gamma }{\delta \bar \psi (x)}
\end{array}
\eqnum{3.13}
\end{equation}
It is obvious that 
\begin{eqnarray}
\frac{\delta W}{\delta u_\mu ^a}=\frac{\delta \Gamma }{\delta u_\mu ^a},\;\;%
\frac{\delta W}{\delta v^a}=\frac{\delta \Gamma }{\delta v^a},\;\;\frac{%
\delta W}{\delta \zeta }=\frac{\delta \Gamma }{\delta \zeta },\;\;\frac{%
\delta W}{\delta \bar \zeta }=\frac{\delta \Gamma }{\delta \bar \zeta } 
\eqnum{3.14}
\end{eqnarray}
Employing Eqs. (3.13) and (3.14), the W-T identity in Eq. (3.10) will be
written as 
\begin{equation}
\begin{array}{c}
\int d^4x\{\frac{\delta \Gamma }{\delta A_\mu ^a(x)}\frac{\delta \Gamma }{%
\delta u^{a\mu }(x)}+\frac{\delta \Gamma }{\delta C^a(x)}\frac{\delta \Gamma 
}{\delta v^a(x)}+\frac{\delta \Gamma }{\delta \psi (x)}\frac{\delta \Gamma }{%
\delta \bar \zeta (x)} \\ 
+\frac{\delta \Gamma }{\delta \bar \psi (x)}\frac{\delta \Gamma }{\delta
\zeta (x)}+\frac 1\alpha \partial _x^\mu A_\mu ^a(x)\frac{\delta \Gamma }{%
\delta \overline{C}^a(x)}\} \\ 
=0
\end{array}
\eqnum{3.15}
\end{equation}
This is the W-T identity satisfied by the generating functional of proper
vertex functions.

The above identity may be represented in another form with the aid of the
so-called ghost equation of motion. The ghost equation may easily be derived
by firstly making the translation transformation: $\bar C^a\rightarrow \bar C%
^a+\bar \lambda ^a$ in Eq.(2.4) where $\bar \lambda ^a$ is an arbitrary
Grassmann variable, then differentiating Eq. (2.4) with respect to the $\bar 
\lambda ^a$ and finally setting $\overline{\lambda }^a=0$. The result is $%
^{[8-11]}$ 
\begin{equation}
\frac 1N\int D(A_\mu ^a,\bar C^a,C^a,\bar \psi ,\psi )\{K^a(x)+\partial
_x^\mu ({\cal D}_\mu ^{ab}(x)C^b(x))\}e^{iS+EST}=0  \eqnum{3.16}
\end{equation}
When we use the generating functional defined in Eq. (3.4) and notice the
relation in Eq. (2.6), the above equation may be represented as$^{[8-11]}$ 
\begin{equation}
\lbrack K^a(x)-i\partial _x^\mu \frac \delta {\delta u^{a\mu }(x)}-i{\mu }^2%
\frac \delta {\delta \bar K^a(x)}]Z[J_\mu ^a,\cdots ,\zeta ]=0  \eqnum{3.17}
\end{equation}
On substituting the relation in Eq. (3.9) into the above equation, we may
write a ghost equation satisfied by the functional $W$ such that 
\begin{equation}
K^a(x)+\partial _x^\mu \frac{\delta W}{\delta u^{a\mu }(x)}+{\mu }^2\frac{%
\delta W}{\delta \bar K^a(x)}=0  \eqnum{3.18}
\end{equation}
From this equation, the ghost equation obeyed by the functional $\Gamma $ is
easy to be derived by virtue of Eqs. (3.12) - (3.14)$^{[8-11]}$ 
\begin{equation}
\frac{\delta \Gamma }{\delta \bar C^a(x)}-\partial _x^\mu \frac{\delta
\Gamma }{\delta u^{a\mu }(x)}-\mu ^2C^a(x)=0  \eqnum{3.19}
\end{equation}
Upon applying the above equation to the last term in Eq. (3.15). the
identity in Eq. (3.15) will be rewritten as 
\begin{equation}
\begin{array}{c}
\int d^4x\{\frac{\delta \Gamma }{\delta A_\mu ^a}\frac{\delta \Gamma }{%
\delta u^{a\mu }}+\frac{\delta \Gamma }{\delta C^a}\frac{\delta \Gamma }{%
\delta v^a}+\frac{\delta \Gamma }{\delta \psi }\frac{\delta \Gamma }{\delta 
\bar \zeta }+\frac{\delta \Gamma }{\delta \bar \psi }\frac{\delta \Gamma }{%
\delta \zeta } \\ 
+m^2\partial ^\nu A_\nu ^aC^a-\frac 1\alpha \partial ^\mu \partial ^\nu
A_\nu ^a\frac{\delta \Gamma }{\delta u^{a\mu }}\} \\ 
=0
\end{array}
\eqnum{3.20}
\end{equation}

Now, let us define a new functional $\hat \Gamma $ in such a manner 
\begin{equation}
\hat \Gamma =\Gamma +\frac 1{2\alpha }\int d^4x(\partial ^\mu A_\mu ^a)^2 
\eqnum{3.21}
\end{equation}
From this definition, it follows that 
\begin{equation}
\frac{\delta \Gamma }{\delta A_\mu ^a}=\frac{\delta \hat \Gamma }{\delta
A_\mu ^a}+\frac 1\alpha \partial ^\mu \partial ^\nu A_\nu ^a  \eqnum{3.22}
\end{equation}
When inserting Eq. (3.21) into Eq. (3.20) and considering the relation in
Eq. (3.22), we arrive at 
\begin{equation}
\begin{array}{c}
\int d^4x\{\frac{\delta \hat \Gamma }{\delta A_\mu ^a}\frac{\delta \hat 
\Gamma }{\delta u^{a\mu }}+\frac{\delta \hat \Gamma }{\delta C^a}\frac{%
\delta \hat \Gamma }{\delta v^a}+\frac{\delta \hat \Gamma }{\delta \psi }%
\frac{\delta \hat \Gamma }{\delta \bar \zeta } \\ 
+\frac{\delta \hat \Gamma }{\delta \bar \psi }\frac{\delta \hat \Gamma }{%
\delta \zeta }+m^2\partial ^\nu A_\nu ^aC^a\} \\ 
=0
\end{array}
\eqnum{3.23}
\end{equation}
The ghost equation represented through the functional ${\hat \Gamma }$ is of
the same form as Eq. (3.19) 
\begin{equation}
\frac{\delta \hat \Gamma }{\delta \bar C^a(x)}-\partial _x^\mu \frac{\delta 
\hat \Gamma }{\delta u^{a\mu }(x)}-{\mu }^2C^a(x)=0  \eqnum{3.24}
\end{equation}
In the Landau gauge, since $\mu =0$ and ${\partial ^\nu A_\nu ^a=0}$, Eqs.
(3.23) and (3.24) respectively reduce to $^{6-8}$ 
\begin{equation}
\int d^4x\{\frac{\delta \hat \Gamma }{\delta A_\mu ^a}\frac{\delta \hat 
\Gamma }{\delta u^{a\mu }}+\frac{\delta \hat \Gamma }{\delta C^a}\frac{%
\delta \hat \Gamma }{\delta v^a}+\frac{\delta \hat \Gamma }{\delta \psi }%
\frac{\delta \hat \Gamma }{\delta \bar \zeta }+\frac{\delta \hat \Gamma }{%
\delta \bar \psi }\frac{\delta \hat \Gamma }{\delta \zeta }\}=0  \eqnum{3.25}
\end{equation}
and 
\begin{equation}
\frac{\delta \hat \Gamma }{\delta \bar C^a}-\partial ^\mu \frac{\delta \hat 
\Gamma }{\delta u^{a\mu }}=0  \eqnum{3.26}
\end{equation}
These equations formally are the same as those for the massless gauge field
theory.

Now, we would like to give another form of the W-T identity. The ghost
equation (3.16) suggests that the first external source term in Eq.(3.5)
which appearing in the generating functional in Eq.(3.4) may be replaced by 
\begin{equation}
\int d^4x\tilde u^{a\mu }(x)\delta {\cal A}_\mu ^a(x)=\int d^4xu^{a\mu
}(x)\Delta {\cal A}_\mu ^a(x)  \eqnum{3.27}
\end{equation}
where 
\begin{equation}
\delta {\cal A}_\mu ^a(x)=\xi {\cal D}_\mu ^{ab}(x)C^b(x)=\xi \Delta {\cal A}%
_\mu ^a(x)  \eqnum{3.28}
\end{equation}
with 
\begin{equation}
\Delta {\cal A}_\mu ^a(x)=\Delta A_\mu ^a+\frac{{\mu }^2}{\Box _x}\partial
_\mu ^xC^a(x)  \eqnum{3.29}
\end{equation}
In this case, from the above relations, we see, the W-T identity in Eq.
(3.8) will be rewritten as 
\begin{equation}
\begin{array}{c}
\int d^4x\{J^{a\mu }(x)\frac \delta {\delta u^{a\mu }(x)}-\bar K^a(x)\frac 
\delta {\delta v^a(x)}-\bar \eta (x)\frac \delta {\delta \bar \zeta (x)}%
+\eta (x)\frac \delta {\delta \zeta (x)} \\ 
-J^{a\mu }(x)\frac{{\mu }^2}{\Box _x}\partial _\mu ^x\frac \delta {\delta 
\bar K^a(x)}+\frac 1\alpha K^a(x)\partial _x^\mu \frac \delta {\delta
J^{a\mu }(x)}\}Z[J_\mu ^a,\cdots ,\zeta ]=0
\end{array}
\eqnum{3.30}
\end{equation}
and the ghost equation in Eq. (3.17) becomes 
\begin{equation}
\{K^a(x)-i\partial _x^\mu \frac \delta {\delta u^{a\mu }(x)}\}Z[J_\mu
^a,\cdots ,\zeta ]=0  \eqnum{3.31}
\end{equation}
Repeating the derivations described In Eqs .(3.9)-(3.15) and (3.17)-(3.24),
one may obtain from Eqs. (3.30) and (3.31) the identities expressed by the
functional ${\hat \Gamma }$%
\begin{equation}
\begin{array}{c}
\int d^4x\{\frac{\delta \hat \Gamma }{\delta A_\mu ^a(x)}\frac{\delta \hat 
\Gamma }{\delta u^{a\mu }(x)}+\frac{\delta \hat \Gamma }{\delta C^a(x)}\frac{%
\delta \hat \Gamma }{\delta v^a(x)}+\frac{\delta \hat \Gamma }{\psi (x)}%
\frac{\delta \hat \Gamma }{\delta \bar \zeta (x)} \\ 
+\frac{\delta \hat \Gamma }{\delta \bar \psi (x)}\frac{\delta \hat \Gamma }{%
\delta \zeta (x)}-\frac{\delta \hat \Gamma }{\delta A_\mu ^a}\frac{\mu ^2}{%
\Box _x}\partial _\mu ^xC^a(x)+m^2\partial _x^\mu A_\mu ^a(x)C^a(x)\} \\ 
=0
\end{array}
\eqnum{3.32}
\end{equation}
\begin{equation}
\frac{\delta \hat \Gamma }{\delta \bar C^a(x)}-\partial _x^\mu \frac{\delta 
\hat \Gamma }{\delta u^{a\mu }(x)}=0  \eqnum{3.33}
\end{equation}
In comparison of Eqs. (3.32) and (3.33) with Eqs. (3.23) and (3.24), we see,
the advantage of using $\delta {\cal A}_\mu ^a$ to define the external
source is that the ghost equation (3.33) becomes homogeneous. However, the
price paying for this advantage is the increase of an inhomogeneous term
(the fifth term) in Eq. (3.32). In the Landau gauge and in the zero-mass
limit, Eqs. (3.32) and (3.33) still reduce to the homogeneous equations
(3.25) and (3.26).

From the W-T identities formulated above, we may derive various W-T
identities obeyed by Green's functions and vertices, as will be illustrated
later on. Particularly, these identities provide a firm basis for the proof
of renormalizability and unitarity problems of the quantum massive gauge
field theory as will be discussed in this paper and the next paper.

\section{Propagators}

In this section, as an application of the W-T identities derived in the
preceding section, we have a particular interest in deriving the W-T
identities satisfied by the massive gluon and ghost particle propagators and
then discussing their renormalization. To derive the mentioned W-T
identities, it is appropriate to start from the W-T identity in Eq. (3.30)
and the ghost equation in Eq. (3.31). While, we would rather here to use the
corresponding identities shown in Eqs. (3.8) and (3.17). Let us perform
differentiations of the identities represented in Eqs. (3.8) and (3.17) with
respect to the external sources $K^a(x)$ and $K^b(y)$ respectively and then
set all the sources except for the source $J_\mu ^a(x)$ to be zero. In this
way, we obtain the following identities 
\begin{eqnarray}
\frac 1\alpha \partial _x^\mu \frac{\delta Z[J]}{\delta J^{a\mu }(x)}+\int
d^4yJ^{b\nu }(y)\frac{\delta ^2Z[J,K,u]}{\delta K^a(x)\delta u^{b\nu }(y)}%
|_{K=u=0}=0  \eqnum{4.1}
\end{eqnarray}
and 
\begin{equation}
\begin{array}{c}
i\partial _\mu ^x\frac{\delta ^2Z[J.K.u]}{\delta u_\mu ^a(x)\delta K^b(y)}%
|_{K=u=0}+i{\mu }^2\frac{\delta ^2Z[J,\bar K,K]}{\delta \bar K^a(x)\delta
K^b(y)}|_{\bar K=K=0} \\ 
+\delta ^{ab}\delta ^4(x-y)Z[J]=0
\end{array}
\eqnum{4.2}
\end{equation}
Furthermore, on differentiating Eq. (4.1) with respect to $J_\nu ^b(y)$ and
then letting the source $J$ vanish, we may get an identity which is, in
operator representation, of the form $^{[8-11]}$%
\begin{equation}
\frac 1\alpha \partial _x^\mu <0^{+}|T[\hat A_\mu ^a(x)\hat A_\nu
^b(y)]|0^{-}>=<0^{+}|T^{*}[\hat {\bar C^a}(x)\hat D_\nu ^{bd}(y)\hat C%
^d(y)]|0^{-}>  \eqnum{4.3}
\end{equation}
where $\hat A_\nu ^a(x)$, $\hat C^a(x)$ and $\hat {\bar C^a}(x)$ stand for
the gauge field and ghost field operators and $T^{*}$ symbolizes the
covariant time-ordering product. When the source $J$ is set to vanish, Eq.
(4.2) will give such an equation$^{[8-11]}$ 
\begin{equation}
\begin{array}{c}
i\partial _y^\nu <0^{+}|T^{*}\{\hat {\bar C^a}(x)\hat D_\nu ^{bd}(y)\hat C%
^d(y)\}|0^{-}> \\ 
+i{\mu }^2<0^{+}|T[\hat {\bar C^a}(x)\hat C^b(y)]|0^{-}>=\delta ^{ab}\delta
^4(x-y)
\end{array}
\eqnum{4.4}
\end{equation}
Upon inserting Eq. (4.3) into Eq. (4.4), we have 
\begin{equation}
\partial _x^\mu \partial _y^\nu D_{\mu \nu }^{ab}(x-y)-\alpha \mu ^2\Delta
^{ab}(x-y)=-\alpha \delta ^{ab}\delta ^4(x-y)  \eqnum{4.5}
\end{equation}
where 
\begin{equation}
iD_{\mu \nu }^{ab}(x-y)=<0^{+}|T\{\hat A_\mu ^a(x)\hat A_\nu ^b(y)\}|0^{-}> 
\eqnum{4.6}
\end{equation}
which is the familiar full gluon propagator and 
\begin{equation}
i\Delta ^{ab}(x-y)=<0^{+}|T\{\hat C^a(x)\hat {\bar C^b}(y)\}|0^{-}> 
\eqnum{4.7}
\end{equation}
which is the full ghost particle propagator. Eq. (4.5) just is the W-T
identity respected by the gluon propagator which establishes a relation
between the longitudinal part of gluon propagator and the ghost particle
propagator. Particularly, in the Landau gauge, Eq. (4.5) reduces to the form
which exhibits the transversity of the gluon propagator. By the Fourier
transformation, Eq. (4.5) will be converted to the form given in the
momentum space as follows 
\begin{equation}
k^\mu k^\nu D_{\mu \nu }^{ab}(k)-\alpha \mu ^2\Delta ^{ab}(k)=-\alpha \delta
^{ab}  \eqnum{4.8}
\end{equation}
The ghost particle propagator may be determined by the ghost equation shown
in Eq. (4.4). However, we would rather here to derive its expression from
the Dyson-Schwinger equation$^{[13]}$ satisfied by the propagator which may
be established by the perturbation method. 
\begin{equation}
\Delta ^{ab}(k)=\Delta _0^{ab}(k)+\Delta _0^{aa^{\prime }}(k){\Omega }%
^{a^{\prime }b^{\prime }}(k)\Delta ^{b^{\prime }b}(k)  \eqnum{4.9}
\end{equation}
where 
\begin{equation}
i\Delta _0^{ab}(k)=i\delta ^{ab}\Delta _0(k)=\frac{-i\delta ^{ab}}{k^2-\mu
^2+i\varepsilon }  \eqnum{4.10}
\end{equation}
is the free ghost particle propagator obtained in paper I and $-i\Omega
^{ab}(k)=-i\delta ^{ab}\Omega (k)$ denotes the proper self-energy operator
of ghost particle. From Eq. (4.9), it is easy to solve that 
\begin{equation}
i\Delta ^{ab}(k)=\frac{-i\delta ^{ab}}{k^2[1+\hat \Omega (k^2)]-\mu
^2+i\varepsilon }  \eqnum{4.11}
\end{equation}
where the self-energy has properly been expressed as 
\begin{equation}
\Omega (k)=k^2\hat \Omega (k^2)  \eqnum{4.12}
\end{equation}
Similarly, we may write a Dyson-Schwinger equation for the gluon propagator
by the perturbation procedure$^{[13]}$ 
\begin{equation}
D_{\mu \nu }(k)=D_{\mu \nu }^0(k)+D_{\mu \lambda }^0(k)\Pi ^{\lambda \rho
}(k)D_{\rho \nu }(k)  \eqnum{4.13}
\end{equation}
where the color indices are suppressed for simplicity and 
\begin{equation}
iD_{\mu \nu }^{(0)ab}(k)=i\delta ^{ab}D_{\mu \nu }^{(0)}(k)=-i\delta ^{ab}[%
\frac{g_{\mu \nu }-k_\mu k_\nu /k^2}{k^2-m^2+i\varepsilon }+\frac{\alpha
k_\mu k_\nu /k^2}{k^2-\mu ^2+i\varepsilon }]  \eqnum{4.14}
\end{equation}
is the free gluon propagator as derived in paper I and $-i\Pi _{\mu \nu
}^{ab}(k)=-i\delta ^{ab}\Pi _{\mu \nu }(k)$ stands for the gluon proper
self-energy operator . Let us decompose the propagator and the self-energy
operator into transverse and longitudinal parts: 
\begin{equation}
D^{\mu \nu }(k)=D_T^{\mu \nu }(k)+D_L^{\mu \nu }(k),\Pi ^{\mu \nu }(k)=\Pi
_T^{\mu \nu }(k)+\Pi _L^{\mu \nu }(k)  \eqnum{4.15}
\end{equation}
where 
\begin{equation}
\begin{array}{c}
D_T^{\mu \nu }(k)=(g^{\mu \nu }-\frac{k^\mu k^\nu }{k^2})D_T(k^2),\text{ }%
D_L^{\mu \nu }(k)=\frac{k^\mu k^\nu }{k^2}D_L(k^2), \\ 
\Pi _T^{\mu \nu }(k)=(g^{\mu \nu }-\frac{k^\mu k^\nu }{k^2})\Pi _T(k^2),%
\text{ }\Pi _L^{\mu \nu }(k)=\frac{k^\mu k^\nu }{k^2}\Pi _L(k^2)
\end{array}
\eqnum{4.16}
\end{equation}
Considering these decompositions and the orthogonality between the
transverse and longitudinal parts, Eq. (4.13) will be split into two
equations 
\begin{equation}
D_{T\mu \nu }(k)=D_{T\mu \nu }^0(k)+D_{T\mu \lambda }^0(k)\Pi _T^{\lambda
\rho }(k)D_{T\rho \nu }(k)  \eqnum{4.17}
\end{equation}
and 
\begin{equation}
D_{L\mu \nu }(k)=D_{L\mu \nu }^0(k)+D_{L\mu \lambda }^0(k)\Pi _L^{\lambda
\rho }(k)D_{L\rho \nu }(k)  \eqnum{4.18}
\end{equation}
Solving the equations (4.17) and (4.18), one can get 
\begin{equation}
iD_{\mu \nu }^{ab}(k)=-i\delta ^{ab}\{\frac{g_{\mu \nu }-k_\mu k_\nu /k^2}{%
k^2+\Pi _T(k^2)-m^2+i\varepsilon }+\frac{\alpha k_\mu k_\nu /k^2}{k^2+\alpha
\Pi _L(k^2)-\mu ^2+i\varepsilon }\}.  \eqnum{4.19}
\end{equation}
With setting

\begin{equation}
\Pi _T(k^2)=k^2\Pi _1(k^2)+m^2\Pi _2(k^2)  \eqnum{4.20}
\end{equation}
which follows from thr Lorentz-covariance of the operator $\Pi _T(k^2)$ and

\begin{equation}
\alpha \Pi _L(k^2)=k^2\hat \Pi _L(k^2),  \eqnum{4.21}
\end{equation}
Eq. (4.19) will be rewitten as 
\begin{equation}
iD_{\mu \nu }^{ab}(k)=-i\delta ^{ab}\{\frac{g_{\mu \nu }-k_\mu k_\nu /k^2}{%
k^2[1+\Pi _1(k^2)]-m^2[1-\Pi _2(k^2)]-+i\varepsilon }+\frac{\alpha k_\mu
k_\nu /k^2}{k^2[1+\hat \Pi _L(k^2)]-\mu ^2+i\varepsilon }.  \eqnum{4.22}
\end{equation}
We would like to note that the expressions given in Eqs. (4.12), (4.20) and
(4.21) can be verified by practical calculations and are important for
renormalizations of the propogators and the gluon mass.

Substitution of Eqs. (4.11) and (4.22) into Eq. (4.8) yields 
\begin{equation}
\hat \Pi _L(k^2)=\frac{\mu ^2\hat \Omega (k^2)}{k^2[1+\hat \Omega (k^2)]} 
\eqnum{4.23}
\end{equation}
From this relation, we see, either in the Landau gauge or in the zero-mass
limit, the $\hat \Pi _L(k^2)$ vanishes.

Now let us discuss the renormalization. The function $\hat \Omega (k^2)$ in
Eq. (4.11), the functions $\Pi _1(k^2)$, $\Pi _2(k^2)$ and $\hat \Pi _L(k^2)$
in Eq. (4.22) are generally divergent in higher order perturbative
calculations. According to the conventional procedure of renormalization,
the divergences included in the functions $\hat \Omega (k^2),$ $\Pi _1(k^2),$
$\Pi _2(k^2)$ and $\hat \Pi _L(k^2)$ may be subtracted at a renormalization
point, say, $k^2=\nu ^2$. Thus, we can write$^{[5-9]}$%
\begin{equation}
\begin{array}{c}
\hat \Omega (k^2)=\hat \Omega (\nu ^2)+\hat \Omega ^c(k^2),\;\;\Pi
_1(k^2)=\Pi _1(\nu ^2)+\Pi _1^c(k^2), \\ 
\Pi _2(k^2)=\Pi _2(\nu ^2)+\Pi _2^c(k^2),\text{ }\hat \Pi _L(k^2)=\hat \Pi
_L(\nu ^2)+\hat \Pi _L^c(k^2)
\end{array}
\eqnum{4.24}
\end{equation}
where $\hat \Omega (\nu ^2)$, $\Pi _1(\nu ^2),$ $\Pi _2(\nu ^2),$ $\hat \Pi
_L(\nu ^2)$ and $\Omega ^c(k^2)$, $\Pi _1^c(k^2)$, $\Pi _2^c(k^2),$ $\hat \Pi
_L^c(k^2)$ are respectively the divergent parts and the finite parts of the
functions $\Omega (k^2)$, $\Pi _1(k^2)$, $\Pi _2(k^2)$ and $\hat \Pi _L(k^2)$%
. The divergent parts can be absorbed in the following renormalization
constants defined by$^{[5-9]}$%
\begin{equation}
\begin{array}{c}
\tilde Z_3^{-1}=1+\hat \Omega (\nu ^2),\;\;Z_3^{-1}=1+\Pi _1(\nu ^2),\text{ }%
Z_3^{\prime -1}=1+\hat \Pi _L(\nu ^2), \\ 
Z_m^{-1}=\sqrt{Z_3[1-\Pi _2(\nu ^2)]}=\sqrt{[1-\Pi _1(\nu ^2)][1-\Pi _2(\nu
^2)]}
\end{array}
\eqnum{4.25}
\end{equation}
where $Z_3$ and $\tilde Z_3$ are the renormalization constants of gluon and
ghost particle propagators respectively, $Z_3^{\prime }$ is the additional
renormalization constant of the longitudinal part of gluon propagator and $%
Z_m$ is the renormalization constant of gluon mass. With the above
definitions of the renormalization constants, on inserting Eq. (4.24) into
Eqs. (4.11) and (4.22) , the ghost particle propagator and gluon propagator
can be renormalized, respectively, in such a manner 
\begin{equation}
i\Delta ^{ab}(k)=\tilde Z_3i\Delta _R^{ab}(k)  \eqnum{4.26}
\end{equation}
and + 
\begin{equation}
iD_{\mu \nu }^{ab}(k)=Z_3iD_{R\mu \nu }^{~~ab}(k)  \eqnum{4.27}
\end{equation}
where 
\begin{equation}
i\Delta _R^{ab}(k)=\frac{-i\delta ^{ab}}{k^2[1+\Omega _R(k^2)]-\mu
_R^2+i\varepsilon }  \eqnum{4.28}
\end{equation}
and 
\begin{equation}
iD_{R\mu \nu }^{ab}(k)=-i\delta ^{ab}\{\frac{g_{\mu \nu }-k_\mu k_\nu /k^2}{%
k^2-m_R^2+\Pi _R^T(k^2)+i\varepsilon }+\frac{Z_3^{\prime }\alpha _Rk_\mu
k_\nu /k^2}{k^2[1+\Pi _R^L(k^2)]-\overline{\mu }_R^2+i\varepsilon }\} 
\eqnum{4.29}
\end{equation}
are the renormalized propagators in which $m_R,$ $\overline{\mu }_R$ and $%
\mu _R$ are the renormalized masses, $\alpha _R$ is the renormalized gauge
parameter, $\Omega _R(k^2),\Pi _R^T(k^2)$ and $\Pi _R^L(k^2)$ denote the
finite corrections coming from the loop diagrams. They are defined as 
\begin{equation}
\begin{array}{c}
m_R=Z_m^{-1}m,\;\alpha _R=Z_3^{-1}\alpha ,\;\overline{\mu }_R=\sqrt{%
Z_3^{^{\prime }}}\mu ,\text{ }\mu _R=\sqrt{\widetilde{Z}_3}\mu , \\ 
\Omega _R(k^2)=\tilde Z_3\hat \Omega ^c(k^2),\text{ }\Pi
_R^T(k^2)=Z_3[k^2\Pi _1^c(k^2)+m^2\Pi _2^c(k^2)],\;\Pi _R^L(k^2)=Z_3^{\prime
}\hat \Pi _L^c(k^2).
\end{array}
\eqnum{4.30}
\end{equation}
The finite corrections above are zero at the renormalization point $\nu $.
As we see from Eq. (4.29), the longitudinal part of the gluon propagator,
except for in the Landau gauge, needs to be renormalized and has an extra
renormalization constant ${Z}_3^{\prime }$. This fact coincides with the
general property of the massive vector boson propagator (see Ref. (8),
Chap.V). From Eqs. (4.23)-(4.25) , it is easy to find that the longitudinal
part in Eq. ( 4.22) can be renormalized as 
\begin{equation}
\frac \alpha {k^2[1+\hat \Pi _L(k^2)]-\mu ^2+i\varepsilon }=Z_3\alpha
_R[1+\Omega _R(k^2)]\Delta _R(k^2)  \eqnum{4.31}
\end{equation}
where 
\begin{equation}
\Delta _R(k^2)=\frac 1{k^2[1+\Omega _R(k^2)]-\mu _R^2+i\varepsilon } 
\eqnum{4.32}
\end{equation}
which appears in Eq. (4.28) and the renormalization constant $Z_3^{\prime }$
can be expressed as 
\begin{equation}
Z_3^{\prime }=[1+\frac{\mu _R^2}{\nu ^2}\frac{(1-\tilde Z_3)}{\tilde Z_3}%
]^{-1}  \eqnum{4.33}
\end{equation}
If choosing $\nu =\mu _R$, we have

\begin{equation}
Z_3^{\prime }=\tilde Z_3  \eqnum{4.34}
\end{equation}

\section{Gauge-independence and unitarity}

This section serves to prove that the S-matrix elements evaluated by the
massive gauge field theory are independent of the gauge parameter. That is
to say, the gauge-dependent spurious pole appearing in the ghost particle
propagator and the longitudinal part of the gauge boson propagator as shown
in Eqs. (4.10) and (4.14) would not contribute to the S-matrix elements.
This fact just ensures the unitarity of the S-matrix. According to the
reduction formula$^{[11]}$, the S-matrix elements can be obtained from the
corresponding Green's functions. So, we first examine how the Green's
functions are dependent on the gauge parameter.

Let us start from the generating functional of Green's functions given in
Eqs. (2.4) and (2.5). Since the fermion field in the generating functional
is not related to the gauge parameter, for simplicity of statement, we will
omit the fermion field functions in the generating functional and rewrite
the functional in the form 
\begin{equation}
\begin{array}{c}
Z[J,\bar K,K]=\frac 1N\int {\cal D}[A.\bar C.C]exp\{iS+i\int d^4x[-\frac 1{%
2\alpha }(\partial ^\mu A_\mu ^a)^2+J^{a\mu }A_\mu ^a \\ 
+\bar K^aC^a+\bar C^aK^a]+i\int d^4xd^4y\bar C^a(x)M^{ab}(x,y)C^b(y)\}
\end{array}
\eqnum{5.1}
\end{equation}
where 
\begin{equation}
S=\int d^4x[-\frac 14F^{a\mu \nu }F_{\mu \nu }^a+\frac 12m^2A^{a\mu }A_\mu
^a]  \eqnum{5.2}
\end{equation}
and 
\begin{equation}
M^{ab}(x,y)=\partial _x^\mu [{\cal D}_\mu ^{ab}(x)\delta ^4(x-y)] 
\eqnum{5.3}
\end{equation}
in which ${\cal D}_\mu ^{ab}(x)$ was defined in Eqs. (2.6) and (2.7).

When we make the following translation transformations in Eq. (5.1) 
\begin{equation}
\begin{array}{c}
C^a(x)\to C^a(x)-\int d^4y(M^{-1})^{ab}(x,y)K^b(y) \\ 
\bar C^a(x)\to \bar C^a(x)-\int d^4y\bar K^b(y)(M^{-1})^{ba}(y,x)
\end{array}
\eqnum{5.4}
\end{equation}
and complete the integration over the ghost field variables, Eq. (5.1) will
be expressed as 
\begin{equation}
Z[J,\bar K,K]=e^{-i\int d^4xd^4y\bar K^a(x)(M^{-1})^{ab}(x,y,\delta /i\delta
J)K^b(y)}Z[J]  \eqnum{5.5}
\end{equation}
where $Z[J]$ is the generating functional without the external sources of
ghost fields$^{[8,12]}$. 
\begin{equation}
Z[J]=\frac 1N\int {\cal D}(A)\Delta _F[A]exp\{iS+i\int d^4x[-\frac 1{2\alpha 
}(\partial ^\mu A_\mu ^a)^2+J^{a\mu }A_\mu ^a]\}  \eqnum{5.6}
\end{equation}
in which 
\begin{equation}
\Delta _F[A]=detM[A]  \eqnum{5.7}
\end{equation}
here the matrix $M[A]$ was defined in Eq. (5.3). From Eq. (5.5), we may
obtain the ghost particle propagator in the presence of the external source $%
J$ 
\begin{equation}
\begin{array}{c}
i\Delta ^{ab}[x,y,J]=\frac{\delta ^2Z[J,\bar K,K]}{\delta \bar K^a(x)\delta
K^b(y)}|_{\overline{K}=K=0} \\ 
=i(M^{-1})_{ab}[x,y,\frac \delta {i\delta J}]Z[J]
\end{array}
\eqnum{5.8}
\end{equation}
The above result allows us to rewrite the W-T identity in Eq. (4.1) in terms
of the generating functional $Z[J]$ when completing the derivative with
respect to $u^{b\nu }(y)$ and setting $K^a(x)=$ $\overline{K}^b(y)=0,$ 
\begin{equation}
\frac 1\alpha \partial _x^\mu \frac{\delta Z[J]}{i\delta J^{a\mu }(x)}-\int
d^4yJ^{b\mu }(y)D_\mu ^{bd}[y,\frac \delta {i\delta J}](M^{-1})^{da}(y,x,%
\frac \delta {i\delta J})Z[J]=0  \eqnum{5.9}
\end{equation}
where 
\begin{equation}
D_\mu ^{bd}(y)={\cal D}_\mu ^{bd}(y)-\frac{\mu ^2}{\Box }\partial _\mu
\delta ^{bd}  \eqnum{5.10}
\end{equation}
is the ordinary covariant derivative. On completing the differentiations
with respect to the source $J$, Eq. (5.9) reads 
\begin{equation}
\begin{array}{c}
\frac 1N\int {\cal D}[A]\Delta _F[A]exp\{iS+i\int d^4x[-\frac 1{2\alpha }%
(\partial ^\mu A_\mu ^a)^2+J^{a\mu }A_\mu ^a]\} \\ 
\times [\int d^4yJ^{b\mu }(y)D_\mu ^{bc}(y)(M^{-1})^{ca}(y,x)-\frac 1\alpha
\partial ^\nu A_\nu ^a(x)] \\ 
=0
\end{array}
\eqnum{5.11}
\end{equation}

By making use of Eqs. (5.3), (5.8) and (5.10), the ghost equation shown in
Eq. (4.2) may be written as 
\begin{equation}
M^{ac}[x,\frac \delta {i\delta J}](M^{-1})^{cb}[x,y,\frac \delta {i\delta J}%
]Z[J]=\delta ^{ab}\delta ^4(x-y)Z[J]  \eqnum{5.12}
\end{equation}
When the source $J$ is turned off, we get 
\begin{equation}
M^{ac}(x)\Delta ^{cb}(x-y)=\delta ^{ab}\delta ^4(x-y)  \eqnum{5.13}
\end{equation}
This equation only affirms the fact that the ghost particle propagator is
the inverse of the matrix $M$.

Now we are in a position to describe the proof of the unitarity mentioned in
the beginning of this section. To do this, it is suitable to use the
generating functional written in Eq. (5.6) and the W-T identity shown in Eq.
(5.11) because the S-matrix only has gluon external lines, without any ghost
particle external line to appear. For simplifying statement of the proof, in
the following, we use the matrix notation to represent the integrals. In
this notation, Eqs. (5.6) and (5.11) are respectively represented as$^{[8]}$ 
\begin{equation}
Z[J]_F=\frac 1N\int {\cal D}(A)\Delta _F[A]e^{i\{S[A]-\frac 12F^2+J\cdot A\}}
\eqnum{5.14}
\end{equation}
and 
\begin{equation}
\frac 1N\int {\cal D}(A)\Delta _F[A]e^{i\{S[A]-\frac 12F^2+J\cdot
A\}}[J_bD_{bc}(M_F^{-1})_{ca}-\frac 1{\sqrt{\alpha }}F_a]=0  \eqnum{5.15}
\end{equation}
where we have defined 
\begin{equation}
F_a\equiv \frac 1{\sqrt{\alpha }}\partial ^\mu A_\mu ^a(x)  \eqnum{5.16}
\end{equation}
with $F$ corresponding to the gauge $\alpha $, the subscript $a,b$ or $c$
stands for the color and/or the Lorentz indices and the space-time variable,
and the repeated indices imply summation and/or integration.

Let us consider the generating functional in the gauge $\alpha +\Delta
\alpha $ where $\Delta \alpha $ is taken to be infinitesimal 
\begin{equation}
Z[J]_{F+\Delta F}=\frac 1N\int {\cal D}(A)\Delta _{F+\Delta F}[A]e^{i\{S[A]-%
\frac 12(F+\Delta F)^2+J\cdot A\}}  \eqnum{5.17}
\end{equation}
In the above, 
\begin{equation}
e^{-\frac i2(F+\Delta F)^2}=e^{-\frac i2F^2}[1+\frac{i\Delta \alpha }{%
2\alpha }F^2]  \eqnum{5.18}
\end{equation}
\begin{equation}
\Delta _{F+\Delta F}[A]=detM_{F+\Delta F}  \eqnum{5.19}
\end{equation}
According to the definitions given in Eqs. (5.3), (2.6) and (2.7), it is
seen that 
\begin{equation}
M_{F+\Delta F}^{ab}=M_F^{ab}+\delta ^{ab}\Delta \alpha m^2  \eqnum{5.20}
\end{equation}
Therefore, 
\begin{equation}
\begin{array}{c}
\Delta _{F+\Delta F}[A]=det[M_F(1+\Delta \alpha m^2M_F^{-1})] \\ 
=detM_Fe^{Trln(1+\Delta \alpha m^2M_F^{-1})} \\ 
=\Delta _F[A][1+\Delta \alpha m^2TrM_F^{-1}]
\end{array}
\eqnum{5.21}
\end{equation}
Upon substituting Eqs. (5.18) and (5.21) into Eq. (5.17), we obtain 
\begin{equation}
\begin{array}{c}
Z_{F+\Delta F}[J]=\frac 1N\int {\cal D}(A)\Delta _F[A]e^{i\{S[A]-\frac i2%
F^2+JA\}} \\ 
\times \{1+\frac{i\Delta \alpha }{2\alpha }F^2+\Delta \alpha m^2TrM_F^{-1}\}
\end{array}
\eqnum{5.22}
\end{equation}
For further derivation, it is necessary to employ the W-T identity described
in Eq. (5.15). Acting on Eq. (5.15) with the operator $\frac 12\Delta \alpha
\alpha ^{-\frac 12}F_a[\frac \delta {i\delta J}]$ and noticing 
\begin{equation}
\begin{array}{c}
iF_\alpha [\frac \delta {i\delta J}]J_be^{iJcAc}=iF_a[\frac \delta {i\delta J%
}]\frac \delta {i\delta A_b}e^{iJ_cA_c}=\frac \delta {\delta A_b}F_a[\frac 
\delta {i\delta J}]e^{iJ_cA_c} \\ 
=e^{iJ\cdot A}\{iJ_bF_a[A]+\frac{\delta F_a[A]}{\delta A_b}\}
\end{array}
\eqnum{5.23}
\end{equation}
we have 
\begin{equation}
\begin{array}{c}
\frac 1N\int {\cal D}(A)\Delta _F[A]e^{i\{S[A]-\frac 12F^2+JA\}}\frac{\Delta
\alpha }{2\sqrt{\alpha }}\{iJ_bD_{bc}[A](M_F^{-1})_{ca}F_a[A] \\ 
+\frac{\delta F_a[A]}{\delta A_b}D_{bc}[A](M_F^{-1})_{ca}-\frac i{\sqrt{%
\alpha }}F^2\}=0
\end{array}
\eqnum{5.24}
\end{equation}
Adding Eq. (5.24) to Eq. (5.22) and considering 
\begin{equation}
\begin{array}{c}
\frac{\delta F_a[A]}{\delta A_b}D_{bc}(M_F^{-1})_{ca}=\frac 1{\sqrt{\alpha }}%
(\partial D)_{ac}(M_F^{-1})_{ca}=\frac 1{\sqrt{\alpha }}[M-\mu
^2]_{ac}(M_F^{-1})_{ca} \\ 
=\frac 1{\sqrt{\alpha }}Tr[1-\mu ^2M_F^{-1}]
\end{array}
\eqnum{5.25}
\end{equation}
where Eqs. (5.10) and (5.3) has been used, one may reach the following
result 
\begin{equation}
\begin{array}{c}
Z_{F+\Delta F}[J]=\frac 1N\int {\cal D}(A)\Delta _F[A]e^{i\{S[A]-\frac 12%
F^2+J\cdot A\}}\{1+i\Delta S+iJ^a[\frac{\Delta \alpha }{2\sqrt{\alpha }}%
D_{ab}(M_F^{-1})_{bc}F_c]\} \\ 
=\frac 1N\int {\cal D}(A)\Delta _F[A]e^{i\{S[A]+\Delta S-\frac 12F^2+J\cdot
A^{\prime }\}}
\end{array}
\eqnum{5.26}
\end{equation}
where 
\begin{equation}
A_a^{\prime }=A_a+\frac{\Delta \alpha }{2\sqrt{\alpha }}%
D_{ab}(M_F^{-1})_{bc}F_c  \eqnum{5.27}
\end{equation}
\begin{equation}
\Delta S=-\frac{i\Delta \alpha }{2\alpha }Tr[1+\mu ^2M_F^{-1}]  \eqnum{5.28}
\end{equation}
in which 
\begin{equation}
TrM_F^{-1}=\int d^4x\Delta ^{aa}(0)=const.  \eqnum{5.29}
\end{equation}
Since the $\Delta S$ is a constant (even though it is infinite), it may be
taken out from the integral sign and put in the normalization constant $N$.
Thus, Eq. (5.26) will finally be represented as 
\begin{equation}
Z_{F+\Delta F}[J]=\frac 1N\int {\cal D}(A)\Delta _F[A]e^{i\{S[A]-\frac 12%
F^2+J\cdot A^{\prime }\}}  \eqnum{5.30}
\end{equation}

In comparison of Eq. (5.30) with Eq. (5.14), it is clear to see that the
difference between the both generating functionals merely comes from the
vector potentials in the external source terms, while, the remaining terms
belonging to the dynamical part in the both generating functionals are
completely the same. This indicates that any change of the gauge parameter
can only lead to different field functions in the source terms of the
generating functional. According to the equivalence theorem, the different
field functions in the source terms does not influence on the S-matrix
elements, it can only affect the renormalization of external lines for the
Green's functions and wave functions$^{[8-12]}$. This point will be
explained more specifically in the following.

The n-point gluon Green's functions computed from the generating functionals 
$Z_F[J]$ and $Z_{F+\Delta F}[J]$ are represented in the position space as 
\begin{equation}
G_F(x_1,x_2,\cdot \cdot \cdot ,x_n)=\langle 0\mid T\{{\bf A}(x_1){\bf A}%
(x_2)\cdot \cdot \cdot {\bf A}(x_n)\}\mid 0\rangle  \eqnum{5.31}
\end{equation}
\begin{equation}
G_{F+\Delta F}(x_1,x_2,\cdot \cdot \cdot ,x_n)=\langle 0\mid T\{{\bf A}%
^{\prime }(x_1){\bf A}^{\prime }(x_2)\cdot \cdot \cdot {\bf A}^{\prime
}(x_n)\}\mid 0\rangle  \eqnum{5.32}
\end{equation}
where ${\bf A}(x_i)$ and ${\bf A}^{\prime }(x_i)$ denote the field operators
corresponding the gauges $F$ and $F+\Delta F$ and $x_i$ designates the
space-time for the i-th particle. Here and afterward, we adopt the matrix
notation to represent the field functions and Green's functions, therefore,
the Lorentz and color indices are suppressed. In light of the
renormalization of the field operators$^{[11]}$ 
\begin{equation}
{\bf A}^{a\mu }(x)=Z_F^{\frac 12}{\bf A}_R^{a\mu }(x)  \eqnum{5.33}
\end{equation}
\begin{equation}
{\bf A}^{\prime }{}^{a\mu }(x)=Z_{F+\Delta F}^{\frac 12}{\bf A}_R^{\prime
a\mu }(x)  \eqnum{5.34}
\end{equation}
where $R$ marks the renormalized quantities and $Z_F$ and $Z_{F+\Delta F}$
are the renomalization constants given in the gauges $F$ and $F+\Delta F$
respectively, the renormalization of the above Green's functions, in the
momentum space, can be written as 
\begin{equation}
G_F(k_{1,}k_2,...,k_n)=Z_F^{\frac n2}G_F^R(k_{1,}k_2,...,k_n)  \eqnum{5.35}
\end{equation}
\begin{equation}
G_{F+\triangle F}(k_{1,}k_2,...,k_n)=Z_{F+\triangle F}^{\frac n2}G_{F+\Delta
F}^R(k_{1,}k_2,...,k_n)  \eqnum{5.36}
\end{equation}
It is well-known that the Green's functions computed from the generating
functionals $Z_F[J]$ and $Z_{F+\Delta F}[J]$ have different external lines,
but the same internal structure. Therefore, by the equivalence theorem and
noticing Eqs. (5.35) and (5.36), we have 
\begin{equation}
\begin{array}{c}
\prod\limits_{i=1}^n\lim_{k_i^2\rightarrow m_R^2}(k_i^2-m_R^2)G_{F+\triangle
F}(k_{1,}k_2,...,k_n) \\ 
=Z_{F+\triangle F}^{\frac n2}/Z_F^{\frac n2}\prod\limits_{i=1}^n\lim_{k_i^2%
\rightarrow m_R^2}(k_i^2-m_R^2)G_F(k_{1,}k_2,...,k_n)
\end{array}
\eqnum{5.37}
\end{equation}
The propagators given in the gauges $F$ and $F+\Delta F$ are respectively
renormalized in such a manner$^{[8]}$

\begin{equation}
D_F(k_i)=Z_FD_R(k_i)=D_F^T(k_i)\widehat{T}+R_F(k_i)  \eqnum{5.38}
\end{equation}
\begin{equation}
D_{F+\triangle F}(k_i)=Z_{F+\triangle F}D_R(k_i)=D_{F+\triangle F}^T(k_i)%
\widehat{T}+R_{F+\triangle F}(k_i)  \eqnum{5.39}
\end{equation}
where $\widehat{T}$ denotes the transverse projector, $D_F^T(k_i)$ and $%
D_{F+\triangle F}^T(k_i)$ come from the transverse parts of the propagators $%
D_F(k_i)$ and $D_{F+\triangle F}(k_i)$ and have a physical pole at $%
k_i^2=m_R^2$ 
\begin{equation}
D_F^T(k_i)=\frac{Z_F}{k_i^2-m_R^2}  \eqnum{5.40}
\end{equation}
\begin{equation}
D_{F+\triangle F}^T(k_i)=\frac{Z_{F+\triangle F}}{k_i^2-m_R^2}  \eqnum{5.41}
\end{equation}
while, $R_F(k_i)$ and $R_{F+\triangle F}(k_i)$ represent the remaining parts
of the propagators $D_F(k_i)$ and $D_{F+\triangle F}(k_i)$ which are regular
at the physical pole, therefore, 
\begin{equation}
\lim_{k_i^2\rightarrow m_R^2}(k_i^2-m_R^2)R_F(k_i)=\lim_{k_i^2\rightarrow
m_R^2}(k_i^2-m_R^2)R_{F+\Delta F}(k_i)=0  \eqnum{5.42}
\end{equation}
According to the reduction formula, the S-matrix elements for multi-gluon
scattering which are evaluated in the gauges $F$ and $F+\Delta F$ may be
respectively represented as 
\begin{equation}
\begin{array}{c}
S_F(k_1,...,k_n)=Z_F^{-\frac n2}\prod\limits_{i=1}^n\lim_{k_i^2\rightarrow
m_R^2}A_0(k_i)(k_i^2-m_R^2)G_F(k_1,...,k_n) \\ 
=Z_F^{\frac n2}\prod\limits_{i=1}^n\lim_{k_i^2\rightarrow
m_R^2}A_0(k_i)D_F^T(k_i)^{-1}G_F(k_1,...,k_n)
\end{array}
\eqnum{5.43}
\end{equation}
\begin{equation}
\begin{array}{c}
S_{F+\triangle F}(k_1,...,k_n)=Z_{F+\triangle F}^{-\frac n2%
}\prod\limits_{i=1}^n\lim_{k_i^2\rightarrow
m_R^2}A_0(k_i)(k_i^2-m_R^2)G_{F+\triangle F}(k_1,...,k_n) \\ 
=Z_{F+\triangle F}^{\frac n2}\prod\limits_{i=1}^n\lim_{k_i^2\rightarrow
m_R^2}A_0(k_i)D_{F+\triangle F}^T(k_i)^{-1}G_{F+\triangle F}(k_1,...,k_n)
\end{array}
\eqnum{5.44}
\end{equation}
where $A_0(k_i)$ is the free wave function of the i-th gluon which
represents the state of transverse polarization and therefore is free of the
gauge parameter. On substituting Eq. (5.37) into Eq. (5.44) and noticing Eq.
(5.43), we find 
\begin{equation}
S_{F+\triangle F}(k_1,...,k_n)=S_F(k_1,...,k_n)  \eqnum{5.45}
\end{equation}
which shows that the S-matrix elements are independent of the gauge
parameter. The gauge-independence of the S-matrix elements implies nothing
but the unitarity of the S-matrix because the gauge-dependent spurious pole
which appears in the longitudinal part of the gluon propagator and the ghost
particle propagator and represent the unphysical excitation of the massive
gauge field in the intermediate states must eventually be cancelled out in
the S-matrix elements. From the construction of the theory, the cancellation
seems to be natural. In fact, in the massive Yang-Mills Lagrangian, all the
unphysical degrees of freedom have been restricted by the constraint
conditions imposed on the gauge field and the gauge group. When these
constraint conditions are incorporated in the Lagrangian, the theoretical
principle we based on would automatically guarantee the cancellation of the
unphysical excitations. This conclusion drawn from the above general proof
can be checked by practical perturbative calculations, as will be
demonstrated in a subsequent paper.

\section{Remarks}

In the last section, we would like to make some remarks on the BRST-
external source terms introduced in Eq. (3.4). Ordinarily, to guarantee the
BRST-invariance of the source terms, the composite field functions $\Delta
\Phi _i$ are required to have the nilpotency property $\delta \Delta \Phi
_i=0$ under the BRST-transformations$^{[8-11]}$. For the massless gauge
field theory, as one knows, the composite field functions are indeed
nilpotent. This nilpotency property is still preserved for the massive gauge
field theory established in the Landau gauge because in this gauge the BRST-
transformations are the same as for the massless theory. However, for the
massive gauge field theory set up in the general gauges, we find $\delta
\Delta \Phi _i$ $\neq 0$, the nilpotency loses, since in these gauges the
ghost field acquires a spurious mass $\mu $. In this case, as pointed out in
section 2, to ensure the BRST-invariance of the source terms, we may simply
require the sources $u_i$ to satisfy the condition denoted in Eq. (3.7). The
definition in Eq. (3.7) for the sources is reasonable. Why say so? Firstly,
we note that the original W-T identity formulated in Eq. (3.2) does not
involve the BRST- sources. This identity is suitable to use in practical
applications. Introduction of the BRST source terms in the generating
functional is only for the purpose of representing the identity in Eq. (3.2)
in a convenient form, namely, to represent the composite field functions in
the identity in terms of the differentials of the generating functional with
respect to the corresponding sources. For this purpose, we may start from
the generating functional defined in Eq. (3.4) to re-derive the identity in
Eq. (3.2). In doing this, it is necessary to require the source terms $%
u_i\triangle \Phi _i$ to be BRST-invariant so as to make the derived
identity coincide with that given in Eq. (3.2). How to ensure the source
terms to be BRST-invariant? If the composite field functions $\triangle \Phi
_i$ are nilpotent under the BRST-transformation, $\delta \Delta \Phi _i=0$,
the BRST-invariance of the source terms is certainly guaranteed.
Nevertheless, the nilpotency of the functions $\triangle \Phi _i$ is not a
uniquely necessary condition to ensure the BRST- invariance of the source
terms, particularly, in the case where the functions $\triangle \Phi _i$ are
not nilpotent. In the latter case, considering that under the BRST-
transformations, the functions $\triangle \Phi _i$ can be, in general,
expressed as $\delta \Delta \Phi _i=\xi \widetilde{\Phi }_i$ where the $%
\widetilde{\Phi }_i$ are some nonvanishing functions, we may alternatively
require the sources $u_i$ to satisfy the condition shown in Eq. (3.7) so as
to guarantee the source terms to be BRST- invariant. Actually, this is a
general trick to make the source terms to be BRST-invariant in spite of
whether the functions $\triangle \Phi _i$ are nilpotent or not. As mentioned
before, the sources themselves have no physical meaning. They are, as a
mathematical tool, introduced into the generating functional just for
performing the differentiations. For this purpose, only a certain algebraic
and analytical properties of the sources are necessarily required.
Particularly, In the differentiations, only the infinitesimal property of
the sources are concerned. Therefore, the sources defined in Eq. (3.7) are
mathematically suitable for the purpose of introducing them. The
reasonability of the arguments stated above for the source terms is
substantiated by the correctness of the W-T identities derived in section 4.
Even though the identities in Eqs. (4.1) and (4.2) are derived from the W-T
identity in Eq. (3.8) which is represented in terms of the differentials
withe respect to the BRST-sources, they give rise to a correct relation
between the gluon propagator and the ghost particle one as shown in Eq.
(4.8). The correctness of the relation in Eq. (4.8) may easily be verified
by the free propagators written in Eqs. (4.10) and (4 14). These propagators
were derived in paper I by employing the perturbation method, without
concerning the BRST-source terms and the nilpotency of the BRST-
transformations. A powerful argument of proving the correctness of the way
of introducing the BRST-sources is that after completing the
differentiations in Eq. (3.8) and setting the BRST-sources to vanish, we
immediately obtain the W-T identity in Eq. (3.2) which is irrelevant to the
BRST-sources. Therefore, all identities or relations derived from the W-T
identity in Eq. (3.8) are completely the same as those derived from the
identity in Eq. (3.2). An important example of showing this point will be
presented in Appendix where the identity in Eq. (5.11) which was derived
from the W-T identities in Eqs. (3.8) and used to prove the unitarity of the
theory can equally be derived from the generating functional in Eq. (5.6)
which does not involve the BRST-sources.

\section{Acknowledgment}

This work is supported by National Natural Science Foundation of China.

\section{Appendix}

To confirm the correctness of the identity given in Eq. (5.11), we derive
the identity newly by starting from the generating functional written in Eq.
(5.6). The generating functional in Eq. (5.6) was directly derived from the
massive Yang-Mills Lagrangian by the Faddeev-Popov method of quantization$%
^{[12]}$. Let us make the ordinary gauge transformation 
\begin{equation}
\delta A_\mu ^a=D_\mu ^{ab}\theta ^b  \eqnum{A1}
\end{equation}
to the generating functional in Eq. (5.6). Considering the gauge-invariance
of the functional integral, the integration measure and the functional $%
\triangle _F[A]=\det M[A],$ we get$^{[8,12]}$%
\begin{equation}
\begin{array}{c}
\delta Z[J]=\frac 1N\int D(A)\triangle _F[A]\int d^4y[J^{b\mu
}(y)+m^2A^{b\mu }(y) \\ 
-\frac 1\alpha \partial ^\nu A_\nu ^b\partial _y^\mu ]D_\mu ^{bc}(y)\theta
^c(y)\exp \{iS+i\int d^4x[-\frac 1{2\alpha }(\partial ^\mu A_\mu
^a)^2+J^{a\mu }A_\mu ^a]\} \\ 
=0
\end{array}
\eqnum{A2}
\end{equation}
According to the well-known procedure, the group parameter $\theta ^a(x)$ in
Eq. (A2) may be determined by the following equation$^{[5,9]}$ 
\begin{equation}
M^{ab}(x)\theta ^b(x)\equiv \partial _x^\mu ({\cal D}_\mu ^{ab}(x)\theta
^b(x))=\lambda ^a(x)  \eqnum{A3}
\end{equation}
where $\lambda ^a(x)$ is an arbitrary function. When setting $\lambda
^a(x)=0,$ Eq. (A3) will be reduced to the constraint condition on the gauge
group (the ghost equation) which is used to determine the $\theta ^a(x)$ as
a functional of the vector potential $A_\mu ^a(x)$. However, when the
constraint condition is incorporated into the action by the Lagrange
undetermined multiplier method to give the ghost term in the generating
functional, the $\theta ^a(x)$ should be treated as arbitrary according to
the spirit of Lagrange multiplier method. That is why we may use Eq. (A3) to
determine the functions $\theta ^a(x)$ in terms of the function $\lambda
^a(x)$ . From Eq. (A3), we solve 
\begin{equation}
\theta ^a(x)=\int d^4x(M^{-1})^{ab}(x-y)\lambda ^b(y)  \eqnum{A4}
\end{equation}
Upon substituting the above expression into Eq. (A2) and then taking
derivative of Eq. (A2) with respect to $\lambda ^a(x),$ we obtain 
\begin{equation}
\begin{array}{c}
\frac 1N\int D(A)\triangle _F[A]\int d^4y[J^{b\mu }(y)+m^2A^{b\mu }(y) \\ 
-\frac 1\alpha \partial _y^\nu A_\nu ^b(y)\partial _y^\mu ]D_\mu
^{bc}(y)(M^{-1})^{ca}(y-x)\exp \{iS+ \\ 
i\int d^4x[-\frac 1{2\alpha }(\partial ^\mu A_\mu ^a)^2+J^{a\mu }A_\mu ^a]\}
\\ 
=0
\end{array}
\eqnum{A5}
\end{equation}
According to the expression denoted in Eq. (2.7) and the identity $%
f^{bcd}A^{c\mu }A_\mu ^d=0$, it is easy to see 
\begin{equation}
A^{b\mu }(y)D_\mu ^{bc}(y)(M^{-1})^{ca}(y-x)=A^{b\mu }(y)\partial _\mu
^y(M^{-1})^{ba}(y-x)  \eqnum{A6}
\end{equation}
By making use of the relation in Eq. (5.10), the definition in Eq. (5.3) and
the equation in Eq. (5.12), we deduce 
\begin{equation}
\begin{array}{c}
\frac 1\alpha \partial _y^\nu A_\nu ^b(y)\partial _y^\mu D_\mu
^{bc}(y)(M^{-1})^{ca}(y-x) \\ 
=\frac 1\alpha \partial ^\nu A_\nu ^b(y)\delta ^4(x-y)-m^2\partial _y^\nu
A_\nu ^b(y)(M^{-1})^{ba}(y-x)
\end{array}
\eqnum{A7}
\end{equation}
On inserting Eqs. (A6) and (A7) into Eq. (A5), we obtain an identity which
is exactly identical to that given in Eq. (5.11) although in the above
derivation, we started from the generating functional without containing the
ghost field functions and the BRST-sources and , therefore, the derivation
does not concern the nilpotency of the composite field functions appearing
in the BRST-source terms. This fact indicates that the W-T identities
derived in section 3 are correct and hence the procedure of introducing the
BRST-invariant source terms into the generating functional is completely
reasonable.

\section{References}

\begin{itemize}
\item[1]  J. C. Su, IL Nuovo Cimento {\bf 117 B} (2002) 203-218.

\item[2]  J. C. Su and J. X. Chen, Phys. Rev.{\bf \ D 69}, 076002 (2004).

\item[3]  J. C. Su, Proceedings of Institute of Mathematics of NAS of
Ukraine, Vol. {\bf 50}, Part 2, 965 (2004).

\item[4]  .J. C. Su and H. J. Wang, Phys. Rev. {\bf C 70}, 044003 (2004).

\item[5]  C. Becchi, A. Rouet and R. Stora, Phys. Lett. {\bf B 52} (1974)
344;

Commun. Math. Phys. {\bf 42} (1975) 127; I. V. Tyutin, Lebedev Preprint {\bf %
39} (1975).
\end{itemize}

\begin{description}
\item[6]  J. C. Ward, Phys. Rev. {\bf 77} (1950) 2931.

\item[7]  Y. Takakashi, Nuovo Cimento {\bf 6} (1957) 370.

\item[8]  E. S. Abers and B. W. Lee, Phys. Rep. {\bf C 9} (1973) 1.

\item[9]  B. W. Lee, in Methods in Field Theory (1975), ed. R .Balian and J.
Zinn-Justin.

\item[10]  W. Marciano and H. Pagels, Phys.Rep {\bf 36} (1978) 137.

\item[11]  C. Itzykson and F-B. Zuber, Quantum Field Theory, McGraw-Hill,
New York (1980).

\item[12]  L. D. Faddeev and A. A. Slavnov, Gauge Fields: Introduction to
Quantum Theory, The Benjamin Commings Publishing Company Inc. (1980).

\item[13]  F. J. Dyson, Phys. Rev. {\bf 75} (1949) 1736; J. Schwinger, Proc.
Nat. Acad. Sci. {\bf 37} (1951) 452.
\end{description}

\end{document}